\documentclass[fleqn]{pasj02} 
\usepackage[switch,mathlines]{lineno} 
\usepackage{mathptmx}
\usepackage{amsmath}
\usepackage{natbib} 
\usepackage{graphicx}
\usepackage{color}

\jyear{2026}
\Received{2026/05/20}
\Accepted{2026/09/30}

\graphicspath{{./}{figures_ando/}} 

\begin{document} 

\title{ A New Analysis Tool for Radial Pulsations }

\author{
 Hiroyasu \textsc{Ando},\altaffilmark{1,2} \email{ando.hr@iris.ocn.ne.jp} 
}
\altaffiltext{1}{National Astronomical Observatory of Japan, National Institutes of Natural Sciences, 2-21-1,Osawa, Mitaka, Tokyo, 181-8588, Japan}
\altaffiltext{2}{Department of Astronomy, Graduate University for Advanced Studies, Shonan Village, Hayama, Kanagawa, 240-0193, Japan}




\maketitle

\begin{abstract}
 Wave-energy relation is introduced as a new analysis tool for linear 
 radial pulsations.  Wave energy consists of kinetic energy, acoustic-potential energy, dynamical-potential energy, and entropy-potential energy. The relation describes changes in the wave energy caused by non-adiabatic processes and by leakage
  of wave energy.    It is applied to modal analysis of adiabatic pulsations for the 
  massive Zero Age Main Sequence
models ($10M_{\Sol}$, $20M_{\Sol}$, and $30M_{\Sol}$). The ordinary modes are  well trapped in the interior cavity without any significant wave-energy leakage.
 On the other hand, the strange modes (S-modes) exhibit  the steep 
 increases in the eigenfunctions and kinetic energy toward the stellar surface 
 and  the significant wave-energy  leakage. The kinetic energy of S-modes is
  no longer confined in the exterior cavity. 
\end{abstract}


\section{Introduction}
Since the discovery of strange modes in radial pulsations by \citet{wood1976}, a tremendous number of theoretical studies on strange modes have been conducted for various types of stars. However, the essential properties of strange modes are not yet 
 completely understood.  For further information and historical details, see  \citet{saio1998}.
 
 \citet{buchler1997} have found that a potential barrier for radial pulsations exists at a dip in the adiabatic exponent $\Gamma_1$ within the hydrogen ionization zone near the surface of Cepheid models.   
Thus, this potential barrier can produce two potential cavities (interior and exterior) for stellar pulsations. They have concluded that ordinary modes are 
 trapped in the interior cavity,  whereas strange modes are ``surface modes'' 
trapped in the exterior cavity, which is like a musical instrument with an open end.
But they did not have discussed wave-energy leakage. It should be noted that these calculations have been done in the adiabatic regime. They also have pointed out that the similar situations
 (existence of barriers) would happen in the nonradial oscillations. In fact, \citet{H&G2007}
 have discussed the effects of abrupt changes of the equilibrium , such as
  $\Gamma_1$ , on frequencies in asteroseismology.
   
 \citet{wood1976}  already noticed that a strange mode has a standing-wave nature in the interior together with a running-wave nature in the exterior, which leads to wave-energy leakage.
 
 Here we introduce a wave-energy relation for radial pulsations, analogous to that for nonradial oscillations presented in Section 25 of \citet{unno1989}.  This relation describes the distribution of wave-energy
  (kinetic and potential energies)  throughout the star, together with its temporal evolution caused by thermal processes and wave-energy leakage. 
  For a deeper understanding of the properties of pulsation modes, modal analysis based on the wave-energy relation is indispensable, along with analyses of eigenvalues and eigenfunctions.    
 In fact, the wave-energy relation for nonradial oscillations is so powerful that \citet{ando2026} has clarified the conditions under which convective modes become oscillatory and the mechanism by which this occurs.
 
 Motivated by these two previous studies, we perform a modal analysis of linear adiabatic pulsations in massive stellar models based on the wave-energy relation.
 In Section 2, we derive the wave-energy relation for radial pulsations from the basic equations of linear radial pulsations and discuss its physical meaning.  
   In Section 3, we calculate adiabatic pulsation modes for three massive  
   stellar models and examine the properties of the pulsation modes in detail 
   based on the wave-energy relation.    
Section 4 is summary and discussion.

\section{Wave-energy relation}\label{sec:2}
\subsection{Basic equation}\label{sec:basic equation}
\indent Following Section 5.3 of \citet{cox1974}, the linearized equations for radial pulsations are written as follows: 
 
 the linearized equation for mass conservation:
\begin{equation}
 \frac{\delta \rho}{\rho} + 3\zeta + r \frac{d\zeta}{dr} =0 ~\   .
 \label{eq:massconv}
\end{equation}
\indent the linearized equation for momentum conservation: 
\begin{equation}
 \frac{d v_r}{dt} - \left(4\zeta + \frac{\delta P}{P}\right)g+ \frac{P}{\rho} \frac{d}{dr}
 \left(\frac{\delta P}{P}\right) = 0 ~\  , 
 \label{eq:momentum}
 \end{equation}
where $\delta$ represents the Lagrangian variation of physical quantities, and $g(=Gm_r/r^2)$, $\zeta=\delta r / r$, and $v_r=d(\delta r)/dt$ denote the local gravitational acceleration, the relative radial displacement, and the radial
 velocity, respectively.
 In order to derive the wave-energy relation, the following thermodynamic relation is also required:
 \begin{equation}
  \frac{\delta \rho}{\rho} = \frac{\delta P}{\varGamma_1 P} - 
  \upsilon_T \frac{\delta S}{C_p} ~\  ,
 \label{eq:therm1}
 \end{equation}
 where $\varGamma_1$ , $\upsilon_T$, and $C_p$ are adiabatic exponent 
 $(\frac{\partial lnp}{\partial ln\rho})_S$, $-(\frac{\partial ln\rho}{\partial lnT})_p$, and specific heat 
 at constant pressure, respectively. 
 
 Following the procedure described in Section 25 of Unno et al. (1989), we take the time derivative of Eq.~(\ref{eq:massconv}), multiply it by $\delta P$, multiply Eq.~(\ref{eq:momentum}) by $\rho v_r$, and add the two resulting equations. After some algebraic manipulations using Eq.~(\ref{eq:therm1}), we obtain
  
 \begin{multline}
\frac{1}{2} \rho \frac{d}{dt} \Biggl\{ v_r^2 +
 \left(\frac{\delta P}{\rho c} \right)^2 - 4rg \zeta^2 \Biggr\}  \\
 +\frac{1}{r^2}
 \frac{d}{dr}\left( r^2 \delta P v_r \right) - \upsilon_T \delta P 
 \frac{d}{dt} \left( \frac{\delta S}{C_p} \right) = 0  ~\  ,
 \label{eq:energy1}
 \end{multline}
where $c$ is sound speed ($c^2= \varGamma_1P / \rho $).

Further, using the thermodynamic relation:
 \begin{equation}
 \nabla_a \frac{\delta P}{P} = \frac{\delta T}{T} 
-  \frac{\delta S}{C_p}  ~\  ,
  \label{eq:therm2}
 \end{equation}
the last entropy term in Eq.~(\ref{eq:energy1}) may be rewritten as,
 \begin{equation}
\upsilon_T \delta P 
 \frac{d}{dt} \left( \frac{\delta S}{C_p} \right) = 
 \rho \delta T  \frac{d \delta S}{dt} - 
 \frac{1}{2} \rho C_p T \frac{d}{dt} 
  \left(\frac{\delta S}{C_p} \right)^2  ~\   ,
  \label{eq:entropy}
 \end{equation}
where  the thermodynamic relation $\upsilon_T P/\nabla_a = \rho C_p T $ is
used.

\noindent Finally, the wave-energy relation Eq.~(\ref{eq:energy1}) is reduced to:
 \begin{equation}
\frac{d}{dt}(\rho e_w)  + \nabla \cdot \bf{\it F}_w  =
  \rho \delta \it T  \frac{d \delta \it S}{dt}  ~\   ,
 \label{eq:energy2}
\end{equation}
where the wave energy per unit mass $e_w$ and 
wave-energy flux $\bf{\it F_w}$ are defined as 

 \begin{equation}
 e_w = \frac{1}{2}  \Biggl\{ v_r^2 +
 \left(\frac{\delta P}{\rho c} \right)^2 - 4rg \zeta^2  +  
 C_p T \left( \frac{\delta S}{C_p} \right)^2 \Biggr\}  ~\ ,
  \label{eq:waveenergy}
 \end{equation}
 and 
 \begin{equation}
 \bf{ \it F} _w= (\it{\delta P v_r, 0, 0})  ~\ ,
  \label{eq:waveflux}
 \end{equation}
 respectively.
 
 \indent Integrating Eq.~(\ref{eq:energy2}) over the entire volume gives 
  \begin{equation}
\frac{d}{dt}(E_w)  + \int_s \bf{\it F}_w \cdot dS =
  \int_v \rho \delta \it T  \frac{d \delta \it S}{dt} dV  ~\   ,
 \label{eq:energy3}
\end{equation}
where $E_w=\int \rho e_w dV$ is the total wave energy.  This equation indicates that the wave energy of a star can change through non-adiabatic processes and wave-energy leakage.   It should therefore be noted that the wave-energy relation (\ref{eq:energy3}) is quite general and provides a useful tool for discussing the overall stability of a star.   

\subsection{Wave energy}\label{sec:wave_energy}
For a deeper physical understanding of pulsation modes, it is necessary to examine not only the eigenfunctions ($\zeta$, $\delta P/P$) and eigenfrequencies, but also the individual energy-density terms appearing in Eq.~(\ref{eq:waveenergy}).  Here
 each term from the left to the right is denoted by $e_k$, $e_p$, $e_{d}$, and $e_s$, respectively,  such as,
 \begin{equation}
 e_k  =   \frac{1}{2}  v_r^2  ~\ ,
 \label{eq:energy_kinetic}
 \end{equation}
\begin{equation}
 e_p  =   \frac{1}{2}  \left( \frac{\delta P}{\varrho c}  \right)^2  ~\  , 
 \label{eq:energy_pressure}
 \end{equation}
\begin{equation}
e_d   =  \frac{1}{2} (- 4rg\zeta ^2) ~\ , 
 \label{eq:energyharmonic}
 \end{equation}
\begin{equation}
e_s  =   \frac{1}{2} C_p T \left( \frac{\delta S}{C_p} \right)^2 ~\ .
  \label{eq:energy_entropy}
 \end{equation}
 
 Obviously, $e_k$ is the kinetic-energy density. The remaining terms represent potential-energy densities associated with the restoring forces. Apparently, $e_p$ is the potential-energy density of an acoustic mode.
  
 $e_d$ is associated with local gravity and is generally important in the stellar core.
  It is always negative and may be associated with secular or dynamical instability 
  (hereafter referred to as the ``dynamical-potential energy'' ).  Therefore, it should be moved to the right-hand side of Eq.~(\ref{eq:energy3}).   This situation is very similar to that of convective modes in nonradial oscillations.
   Thus, an oscillatory secular mode may arise when the dynamical timescale becomes comparable to the thermal timescale, as suggested by \citet{ando2026} in his study of oscillatory convection.
 
 $e_s$ is a unique energy density, hereafter referred to as the
  ``entropy-potential energy'' which provides a restoring force when the thermal 
  timescale of non-adiabatic processes becomes comparable to the wave timescale,
   as demonstrated by \citet{ando2026}. 
  
  Within the adiabatic framework considered here, it should be noted that only acoustic modes appear, although the effect of the dynamical potential-energy density $e_d$ remains.    

%
\section{Adiabatic analysis of pulsation modes}
\subsection{Linear adiabatic pulsations}
Assuming that the Lagrangian perturbations vary as $e^{i\sigma t}$, we have
 $v_r= i\sigma \delta r $.
Under the adiabatic condition ($\delta S = 0$), the basic equations for adiabatic pulsations are derived from Eqs.~(\ref{eq:massconv}), (\ref{eq:momentum}), and (\ref{eq:therm1}) as follows:
\begin{equation}
 x \frac{dy_1}{dx} = -3y_1 - \frac{1}{\varGamma_1} y_2  ~\ ,
 \label{eq:ada1}
 \end{equation}
\begin{equation}
 x \frac{dy_2}{dx}  =  ( 4 + C_1 \omega ^2 ) V y_1 + Vy_2   ~\  , 
 \label{eq:adia2}
 \end{equation}
where $x= r/R$, $y_1= \zeta$, $y_2= \delta P/ P$, $V= - (dlnP/dlnr)$, 
$C_1 = x^3/(m_r/M)$, and $\sigma ^2 = \omega ^2 (GM/R^3)$.

The above equations are numerically solved under the following boundary conditions:

the inner mechanical boundary condition at $x=0$:
\begin{equation}
 3y_1 + \frac{1}{\varGamma_1} y_2  = 0  ~\  , 
 \label{eq:cond1}
 \end{equation} 
  
the normalization at $x=0$:
\begin{equation}
 y_1   = 1   ~\  , 
 \label{eq:cond2}
 \end{equation} 
 
the outer mechanical boundary condition at $x=1$:
\begin{equation}
  (4 + C_1 \omega ^2 ) y_1 + y_2  = 0 ~\  , 
 \label{eq:cond3}
 \end{equation} 
 
\subsection{Massive ZAMS models} 
As equilibrium models, Zero Age Main Sequence (ZAMS) models with
$M/M_{\Sol} = 10.0, ~20.0, ~30.0$ are calculated by \citet{pacz1969} program,  parameters of which 
  are the ratio of the mixing length to the pressure scale 
 height $l/H_p= 1.5$  and chemical composition $X= 0.7$, $Z= 0.03$. 
 Their parameters of these models are summarized in Table \ref{tab:stellar-para}.
 
\begin{table}[!h]
\caption{ Stellar parameters for massive ZAMS models
}
\label{tab:stellar-para}
\scalebox{1.2}[1.2]{
\begin{tabular}{lcc}
\hline
$M_{\Sol}$ & $log(T_e)$ & $log(L_{\Sol})$   \\
\hline
\hline
10.0 & 4.395 & 3.739 \\
20.0 & 4.533 & 4.612 \\
30.0 & 4.597 & 5.045 \\
\hline
\end {tabular}
}
\end{table}

\begin{figure}[!h]
    \includegraphics[width=0.8\columnwidth,clip]{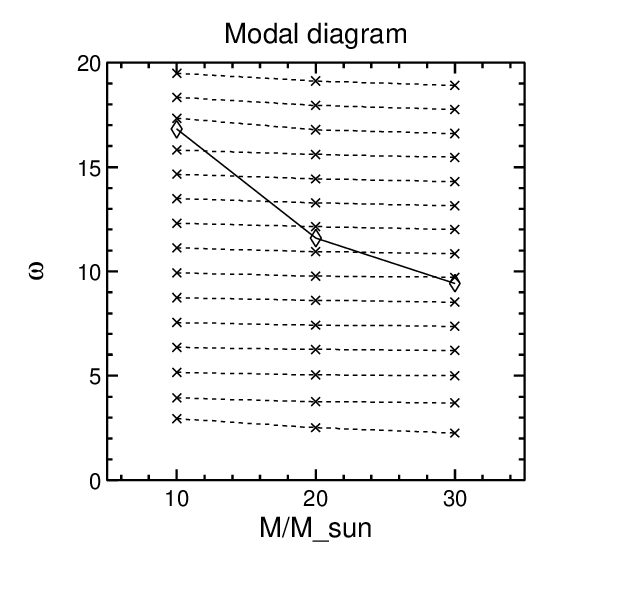}
    \caption{Modal diagram for massive ZAMS models.   \\
    Modes marked with (x) are ordinary modes. Modes marked with ($\diamond$) are
    S-modes.  \\
    {Alt text: x axis shows stellar mass (solar mass unit).  \\
     y axis shows non-dimensional frequencies of pulsation modes. }
      }
    \label{fig:modal_fig}
\end{figure}

Pulsation modes from the fundamental mode to the 14th overtone are calculated for these models, and the results are summarized in the modal diagram (Fig.~\ref{fig:modal_fig}).
In this figure, modes marked with (x) are ordinary modes, whereas those marked with ($\diamond$) are strange modes, hereafter abbreviated as ``S-modes''.
 
\subsection{Analysis of modes} 
\begin{figure}[!h] 
       \includegraphics[width=0.95\columnwidth,clip]{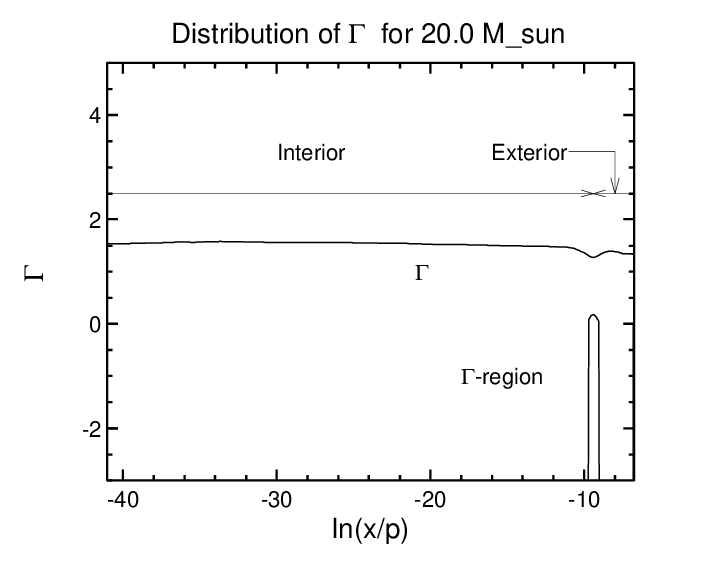} 
    \caption{The distribution of $\varGamma_1$ exhibits a dip in the He ionization zone together with the $\varGamma$-region (i.e., where $(3\varGamma_1 - 4) < 0$).
    The correspondence of a dip and $\varGamma$-region  
    is quite good.  Two potential cavities are also indicated. \\
    {Alt text: the distribution of $\varGamma_1$ is given. x-axis is an indicator of stellar
    radius. }
     }
    \label{fig:gamma}
\end{figure}

To clarify the properties of the pulsation modes, we focus on a detailed analysis of
modes for the 20~$M_{\odot}$ model using the wave-energy relation. Fig.~\ref{fig:gamma} displays the distribution of $\varGamma_1$ together with the region where $3\varGamma_1 - 4 < 0$ (denoted as the $\varGamma$-region).
  It is striking that the dip in $\varGamma_1$ produced by the He ionization zone corresponds closely to the $\varGamma$-region. The barrier pointed out by \citet{buchler1997} is thus confirmed to exist very near the surface.
 
%
 
To begin the analysis based on the wave-energy relation, we define the integrated energies $E_k$, $E_p$, $E_d$, and $E_s$ as the volume integrals of the respective energy densities (i.e., $E_k = \int \rho e_k dV$, etc.).  
 The leakage of wave energy, $E_f$, is estimated from Eq.~(\ref{eq:waveflux})
 as follows:
 
 \begin{align}
 \int_S  \bf{\it F}_w \cdot \bf{\it dS}    
&=  4\pi R^2 \delta P v_r   \notag \\
          &= - \frac{1}{2} \frac{d}{dt} \Biggl\{ 4\pi R^3 P (4 + C_1 \omega ^2) \zeta^2 \Biggr\}  \notag  \\
          &= \frac{1}{2} \frac{d}{dt} E_f  ~\  ,
 \label{eq:flux}
 \end{align} 
 where the outer boundary condition Eq.~(\ref{eq:cond3}) is used.

\begin{figure}[!h]
	\includegraphics[width=1.05\columnwidth,clip]{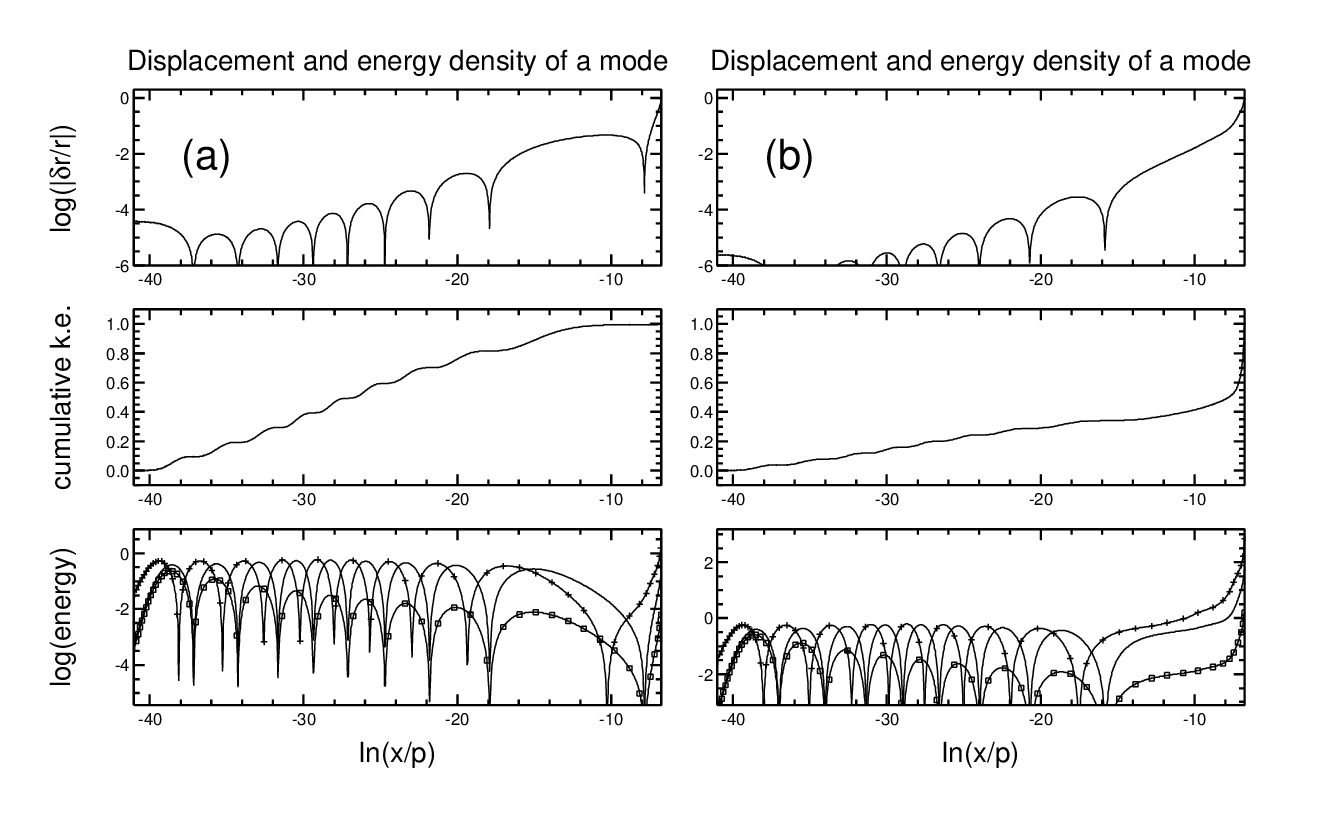} 
    \caption{(a) O8-mode:  Displacement $\zeta$ (top), cumulative kinetic \\
       energy (middle), and 
       distributions  of $e_k$(solid line),\\
        $e_p$(solid line marked with (+)), and $e_d$
    (solid line marked with $\square$) (bottom) \\
    (b) S-mode:  The same presentations as those of  O8-mode. \\
   note: each energy distribution is given per unit spherical shell mass.
    \newline 
    
    {Alt text: Two line graphs. These graphs show the behaviors of \\
     S-mode and ordinary mode O8, respectively. x axis is an indicator of stellar
    radius. } 
     }
    \label{fig:behavior}
\end{figure}

  The results for the modes from F to O9, including the S-mode, are summarized in Table~\ref{tab:mode} and Fig.~\ref{fig:behavior}.   Here, the energy densities
   $e_k$, $e_p$, and $e_d$ are evaluated as time averages over one oscillation period.  The energy distributions shown in Fig.~\ref{fig:behavior} are expressed
  per unit spherical shell mass.
   All integrated energies ( $E_k$, $E_p$, $E_d$, and $E_f$) are normalized by
    $E_k$ (i.e., $E_k = 1$).  $E_{\sum}$ is
  the summation of $E_p$ and $E_d$ (i.e., total potential energy).   
  The cumulative kinetic energy $E_{\varGamma}$ at the barrier is also presented.

\begin{table}[!h]
	\caption{Characteristics of modes for $20M_{\Sol}$ ZAMS. \\
        $E_{\sum}$ is summation of $E_p$ and $E_d$. $E_{\varGamma}$ is 
        cumulative kinetic energy at the barrier.
	}
	\label{tab:mode}
	\scalebox{0.93}[0.93]{
	\begin{tabular}{lccccccr} 
		\hline
		Mode   & $\omega$ & $E_k$ &  $E_p$ & $E_d$ & $E_{\sum}$  & $E_f$ & $E_{\varGamma}$  \\
		\hline
		\hline
		 F   &  2.51 & 1.000 & 6.289 & -5.271 &  1.018 &  0.000 & 1.000 \\
		 O1 & 3.76 & 1.000 & 2.235 & -1.234 & 1.001  &  0.000  & 1.000 \\
		 O2 & 5.05 & 1.000 & 1.662 & -0.662 & 1.000 &  0.000 & 1.000 \\
		 O3 & 6.26 & 1.000 & 1.437 & -0.436 & 1.001 & 0.000  & 1.000 \\
		 O4 & 7.43 & 1.000 & 1.317 & -0.316 & 1.001 &  0.000 &  1.000 \\
		 O5 & 8.60 & 1.000 & 1.237 & -0.238 & 0.999  & 0.000  &  1.000 \\
		 O6 & 9.77 & 1.000 & 1.187 & -0.188 & 0.999 &  0.000 &  0.999 \\
		 O7 & 10.9 & 1.000 & 1.167 & -0.155 & 1.012 &  -0.013 & 0.995 \\
		 S   & 11.6 & 1.000 & 5.319 & -0.073 & 5.246 &  -4.119 & 0.432 \\
		 O8 & 12.1 & 1.000 & 1.179 & -0.130 &  1.049 &  -0.048 & 0.995 \\
		 O9 & 13.3 & 1.000 & 1.116 & -0.109 & 1.007  &  -0.008 &  0.996 \\
		\hline
	\end{tabular}
	}
\end{table}

Now, we discuss the characteristics of modes following  arguments in the Subsection 15.3 of \citet{unno1989}. If a star has multiple cavities in the equilibrium, each cavity has its own intrinsic spectrum of p- or g-modes. 
If the frequency of a mode in one cavity happens to be very close to that of a mode in another cavity, an avoided crossing may occur.  A key issue in mode classification is identifying the cavity in which most of the kinetic energy of a mode is concentrated. 

\noindent (a): Ordinary mode  \\
  As shown in Table~\ref{tab:mode},  cumulative kinetic energy $E_{\varGamma}$ at the barrier is almost equal to
  $E_k (=1)$ in the ordinary modes of F, O1-O9.   Therefore,
 the ordinary modes are considered to be eigenmodes in the interior cavity.
 The total potential energy $E_{\sum}$  is nearly equal to the kinetic energy 
$E_k$, and thus the equipartition principle is satisfied for ordinary modes. 
 It should be noted here that the F- and O1-modes, with periods close to the dynamical timescale, exhibit a significant dynamical-potential energy $E_d$.  However, the sum of
  pressure-potential energy $E_p$ and  $E_d$  is nearly equal to $E_k$.  It stands for still
   maintaining the equipartition principle. 
 
 In Fig.\ref{fig:behavior} (a), the behavior of O8-mode is displayed.
 The top panel shows the relative displacement, $\textrm{log}(|\delta r / r|)$,
where the positions of the nodes are clearly visible.  In the middle panel, the cumulative kinetic energy shows how the local region contributes to the kinetic energy of a mode. 
 In the bottom panel, the distributions of kinetic energy $e_k$ (solid line),
  pressure-potential energy $e_p$ (solid line marked with (+)), and dynamical-potential 
 energy $e_d$ (line marked with ($\square$)) are shown. 
 Here it should be reminded that each energy distribution is given per unit spherical shell mass. It is interesting to note that the kinetic energy $e_k$ (solid line in the bottom panel) and the pressure-potential energy $e_p$ (solid line with (+)) are
 uniformly distributed from the stellar core to the barrier.  The phase 
between $e_k$ and $e_p$ is shifted by 90 degrees, indicating that the equipartition principle is locally satisfied as well.  

\noindent
(b): S-mode  \\  
S-mode is accompanied by significant wave-energy leakage $E_f=4.119$, as shown in Table \ref{tab:mode}.  In adiabatic regime
($\delta S = 0$), Eq.(\ref{eq:energy3}) states that time variation of wave energy
and divergence or convergence of wave energy should be balanced.  Thus,
 this extra wave energy $E_f$ should be considered mathematically to be supplied
  from the outside to keep a complete wave cycle.   Physically it means 
 wave-energy leakage.  
The behavior of S-mode is  shown in Fig.\ref{fig:behavior} (b).  
The steep increases in the eigenfunctions and kinetic energy toward the stellar surface indicate that the exterior cavity is open to the outside.  This is consistent with the physical picture proposed by \citet{buchler1997}, in which the physical size of the exterior cavity in Cepheid models is approximately equal to one quarter of the wavelength.  In fact, the modes with frequencies above
 the critical frequency $\omega_{cr}$ (2.23 for $20M_{\Sol}$) defined by \citet{ledoux1958} may have a running wave nature.   

In S-mode,  56.8\% of the kinetic energy $E_k$ is confined in the very narrow 
exterior cavity, while 43.2\% of  
 $E_k$ remains in the interior cavity as shown in column of  $E_{\varGamma}$ 
 in Table \ref{tab:mode} .  In this point, our situation is different from that of the 
 perfect reflection at the surface discussed in Subsection15.3 by \citet{unno1989}.
 The kinetic energy no longer serves as the primary criterion for oscillation-mode classification.
 In fact, from the summary of the characteristics of S-modes for three ZAMS models
 in Table \ref{tab:mode2}, S-mode for $10M_{\Sol}$ has only 17.7\% of the 
 total kinetic energy as the kinetic energy in the exterior cavity.   
 Therefore, S-mode may be considered to be accompanied by  the steep 
 increases in the eigenfunctions and kinetic energy toward the stellar surface 
 and  the significant wave-energy  leakage.  Of course, it is necessary for the
 exterior cavity to have a significant fraction of the kinetic energy.

\begin{table}[!h]
	\caption{Characteristics of S-modes  for three  ZAMS models \\
	($10M_{\Sol}$, $20M_{\Sol}$, $30M_{\Sol}$). \\
        $E_{\sum}$ means summation of $E_p$ and $E_d$. 
	}
	\label{tab:mode2}
	\scalebox{0.90}[0.90]{
	\begin{tabular}{lccccccr} 
		\hline
		
		Mode   & $\omega$ & $E_k$ &  $E_p$ & $E_d$ & $E_{\sum}$ & $E_f$ & 
		$E_{\varGamma}$ \\
		\hline
		\hline
		         &  &  & $10M_{\Sol}$ &  &  &    \\
		 S-mode   &  16.8 & 1.000 & 1.444 & -0.071 &  1.373 &  -0.368 & 0.823 \\
		 \hline
		        &  &  & $20M_{\Sol}$ &  &  &  \\
		 S-mode & 11.6 & 1.000 & 5.319 & -0.073 &  5.246 &  -4.119 & 0.432 \\
		 \hline
		        &  &  & $30M_{\Sol}$ &  &  &  \\
		 S-mode & 9.44 & 1.000 & 6.813 & -0.091 &  6.722 &  -5.502 &  0.405 \\
		\hline
	\end{tabular}
	}
\end{table}
\indent
We would like to add one further point: the leakage of wave energy is compensated near the surface by an excess of pressure-potential energy $e_p$ over $e_k$, as shown in the bottom panel of Fig.~\ref{fig:behavior}(b).  
 The sum ($E_{\sum} + E_f$= 1.127)
 may suggest that, even in this situation, the star tends to maintain the equipartition principle.   

\section{Summary and discussion}
 The wave-energy relation has proven to be a powerful tool for analyzing modes of radial pulsation. Wave energy consists of kinetic energy, 
acoustic-potential energy, dynamical-potential energy, and entropy-potential energy.
The three potential energies may correspond to the acoustic, secular (or dynamically unstable), and thermal modes, respectively. 
\\
 \indent We  calculate modes from the fundamental mode to the 14th overtone of adiabatic radial pulsations for the three massive
  ZAMS models (10$M_{\Sol}$, 20$M_{\Sol}$, and 30$M_{\Sol}$) and present the modal diagram.   
\\
\indent As a representative example, a detailed modal analysis for 20$M_{\Sol}$
 ZAMS model is performed using the wave-energy relation.
 We confirm the existence of a potential barrier at a dip in
 $\varGamma_1$ associated with the He ionization zone, as pointed out by \citet{buchler1997}.
 This barrier gives rise to two cavities (an interior cavity and an exterior cavity).
\\
\indent  For ordinary modes (the fundamental and overtone modes), the kinetic energy is well confined within the interior cavity, and the equipartition principle is satisfied both globally and locally without any wave-energy leakage.
\\
\indent  
The S-mode exhibits the steep 
 increases in the eigenfunctions and kinetic energy toward the stellar surface 
 and  the significant wave-energy  leakage.  The distribution of kinetic energy
 is no longer confined in the exterior cavity. This situation is different from that of
 ordinary modes. Here the further analysis may be required for this situation.
\citet{P&T2022} have done the detailed analysis of elementary processes of 
 interference of progressive waves in multiple cavities with the perfect reflection surface to prove the argument in the Subsection 15.3 of \citet{unno1989}.
 We expect such a work would be extended to the case of the open
 surface.    
\\
\indent Since stars have several dips in the adiabatic 
exponent $\Gamma_1$ , S-mode is expected to be a ubiquitous 
phenomenon in stars.  
 In fact, our preliminary calculation for a slightly evolved $2M_{\Sol}$
main-sequence model demonstrates the existence of the S-mode. 
At the moment, no S-mode is found for the $2M_{\Sol}$ ZAMS model and for the present Sun. However, intensive search might make it possible to discover
S-modes. 

 We believe that the discussion of S-modes by \citet{saio1998} may represent the case of multiple barriers.  Although their situation is different from that discussed here, a new analysis tool based on the wave-energy relation could 
 be still useful. 
 
\section*{Acknowledgements}
I would like to thank the referee for improving the paper with their helpful comments. 



\bibliographystyle{mnras}    
\bibliography{ando} 


\end{document}